\documentclass[10pt,journal]{IEEEtran}
\usepackage{cite}
\usepackage{graphicx}
\usepackage{subfigure}
\usepackage{algorithm} 
\usepackage{algorithmic}
\usepackage{amsfonts}
\usepackage{amsmath}
\usepackage{subeqnarray}
\usepackage{cases}
\usepackage{flushend}
\usepackage[outdir=./]{epstopdf}

\begin{document}
%
\title{Energy Aware Trajectory Optimization for Aerial Base Stations}
%
%
%

\author{Jingcong~Sun,~\IEEEmembership{Student Member,~IEEE,}
        Christos~Masouros,~\IEEEmembership{Senior Member,~IEEE}\thanks{J.Sun and C.Masouros are with the Department of Electronic and Electrical
Engineering,  University  College  London,  London  WC1E  7JE,  U.K.  (e-mail: uceejsu@ucl.ac.uk; chris.masouros@ieee.org). This work was supported by the Engineering and Physical Science Research Council, U.K., under project EP/R007934/1}}

\maketitle

\begin{abstract}
By fully exploiting the mobility of unmanned aerial vehicles (UAVs), UAV-based aerial base stations (BSs) can move closer to ground users to achieve better communication conditions. In this paper, we consider a scenario where an aerial BS is dispatched for covering a maximum number of ground users before exhausting its on-board energy resources. The resulting trajectory optimization problem is a mixed integer non-linear problem (MINLP) which is non-convex and is challenging to solve. As such, we propose an iterative algorithm which decomposes the problem into two sub-problems by applying both successive convex optimization and block coordinate descent techniques to solve it. To be specific, the trajectory of the aerial BS and the user scheduling and association are alternately optimized within each iteration. In addition, to achieve better coverage performance and speed up convergence, the problem of designing the initial trajectory of the UAV is also considered. Finally, to address the unavailability of accurate user location information (ULI) in practice, two different robust techniques are proposed to compensate the performance loss in the existence of inaccurate ULI. Simulation results show both energy and coverage performance gains for the proposed schemes compared to the benchmark techniques, with an up to 50\% increase in coverage probability and an up to 20\% reduction in energy.
\end{abstract}

\begin{IEEEkeywords}
Unmanned aerial vehicles, aerial base stations, trajectory optimization, user scheduling and association
\end{IEEEkeywords}

%
\IEEEpeerreviewmaketitle

\section{Introduction}
%
%
%
%
\IEEEPARstart{U}{nmaned} aerial vehicles (UAVs) have the ability to provide reliable wireless communication solutions for a wide range of real-world scenarios, and are thus gaining significant popularity in both industrial and academic research. In particular, UAVs serving as aerial base stations (BSs) have increasingly been the focus of wireless service providers, thanks to their flexibility, interoperability and the favourable line-of-sight (LoS) communication conditions \cite{DBLP:journals/corr/abs-1803-00680,7470933}. Aerial BSs can be deployed to ease the burden of existing cellular systems in extremely crowded areas \cite{8292783,7461487}. Moreover, the deployment of aerial BSs is also relevant in emergency or disaster scenarios where ground communication infrastructures are damaged or even totally destroyed \cite{7470933, Namuduri:2013:MAH:2491260.2491265}.\\
\indent One line of research is focused on the deployment of static aerial BSs. The authors in \cite{6863654} first studied the relationship between the altitude of static aerial BS and the corresponding coverage area on the ground. The work in \cite{7417609} then extended the number of deployed UAVs to two by considering the effect of inter-cell interference (ICI). In addition, various algorithms have been proposed to maximize the number of users that can be covered by the static aerial BSs \cite{7918510,7510820,8587183}. Specifically, in \cite{7918510}, a 3-D circle placement problem was formulated as a mixed integer non-linear problem (MINLP) to maximize the ground user coverage probability. In \cite{8587183}, further steps were made to study the coverage-efficient and energy-efficient deployment of multiple aerial BSs by leveraging geometrical relaxation and clustering methods. \\
\indent In order to fully exploit the potential of aerial BSs, recent focus has shifted towards the high mobility of UAVs. With the exploitation of controllable UAV mobility, communication path loss can be greatly reduced due to the reduced distance between the aerial BS and users \cite{8329013}. By assuming the users are distributed along a one-dimensional line, a novel cyclical multiple access (CMA) method was proposed for moving aerial BSs in \cite{7556368}. Based on the access model in \cite{7556368}, authors in \cite{8304088} and \cite{8247211} further proved that increased maximum throughput gain can be obtained by exploiting the UAV mobility for delay-tolerant applications. It is worth mentioning that the endurance of aerial BSs is fundamentally constrained by the limited built-in battery energy, and the efficient use of on-board energy is thus of paramount importance in UAV related applications \cite{7470933,8269064}. Without considering the propulsion energy for supporting the movement of UAVs, efficient usage of energy for communication related functions have been studied in \cite{7510870,7192644,6978873}. Authors in \cite{7888557} took into account the propulsion power consumption and gave the expression of propulsion power with regard to velocity and acceleration. Furthermore, the total power consumption of a UAV was minimized in \cite{8329973} with a guaranteed transmission rate. Nevertheless, the bottom line aim of UAV application is to maximize coverage with a given energy budget. Nevertheless, maximizing the number of covered ground users with a moving aerial BS is not considered yet. On one hand, compared to static aerial BSs, moving aerial BSs can fly close to ground users to improve the channel quality, thus serving more users during a specific time period. On the other hand, the trajectory of UAV is intrinsically constrained by the limited on-board energy which becomes an obstruction for serving more users. \\
\indent In this paper, we consider a scenario that an aerial BS is dispatched from base to meet the service requirement of delay-tolerant ground users. During a given mission period, the aerial BS is required to fly back to the initial position for recharging before exhausting its on-board energy. We assume that a user is covered only when the entire data requested is delivered. Following \cite{7486987,7762053}, we further assume that the user location information (ULI) is known by the aerial BS with the assistance of high-accuracy GPS systems. Fixed-wing UAVs which have higher speed than roatry-wing UAVs are chosen as the carrier for aerial BSs \cite{7470933}. Our aim is to maximize the number of covered ground users with a limited on-board energy resource by jointly optimizing the UAV trajectory and the user communication scheduling and association. However, such a joint optimization problem is not only non-convex but also contains integer variables, and is thus challenging to solve. Successive convex optimization and block coordinate descent method are applied for solving this specific problem. In addition, to improve the convergence behavior and coverage performance, a new initial trajectory is designed for the iterative algorithm. Finally, as the ULI will not be perfectly accurate in practice, two different robust techniques are further proposed to compensate for the performance loss in the existence of inaccurate ULI.\\
\indent For clarity, the main contributions of this paper are summarized as follows
\begin{itemize}
    \item First, we formulate the optimization problem for achieving the maximum number of covered users while considering constraints on energy resources and flying status. We solve this problem by applying successive convex optimization and block coordinate descent techniques. Specifically, the entire set of optimization variables are divided into two subsets: one containing user scheduling and association variables and one consisting of variables related to the mobility of the UAV. The two subsets of variables are alternately optimized within each iteration. However, even by fixing the association variables, the UAV path optimization problem is still challenging to solve due to the non-convex constraints. Successive convex optimization technique is then utilized to find the convex lower-bound of the specific non-convex constraints to solve the problem.
    \item Next, since the convergence and performance of such iterative algorithms depends on the adopted initial trajectory fed to the optimization algorithm  \cite{7366709,8443133}, we consider the design of the initial trajectory. We adopt an initial trajectory which connects all the ground user locations to give all users a fair chance to be scheduled and associated. Simulation results demonstrate that the designed initial trajectory not only speeds up the convergence but also increases the number of covered users.
    \item Furthermore, the existence of imperfect ULI is also considered. For compensating the resulting performance loss, two robust techniques are proposed. Specifically, instead of considering the user location acquired by the GPS, the first robust technique optimizes the variables with respect to the worst-case user location which leads to the highest path loss at each time slot. Since a user is covered only when the data transmitted is equal or larger than the demand requested, the second robust technique increases robustness by maximizing the excess between provided bits and the requested demand. Such an optimization problem is non-convex, but can again be efficiently solved by applying successive convex optimization and block coordinate descent techniques. 
\end{itemize}

\indent The remainder of this paper is organized as follows. Section \uppercase\expandafter{\romannumeral2} introduces system model and formulates the problem. The proposed iterative algorithm for solving the optimization problem is shown in Section \uppercase\expandafter{\romannumeral3}. Section \uppercase\expandafter{\romannumeral4} designs an initial trajectory for achieving faster convergence and better serving performance. Existence of inaccurate ULI is considered in Section \uppercase\expandafter{\romannumeral5}, and two different robust techniques are proposed successively in this section. In Section \uppercase\expandafter{\romannumeral6}, benefits of the proposed techniques are evaluated with numerical results. Finally, the paper is concluded in Section \uppercase\expandafter{\romannumeral7}.



\begin{figure}
\centering
\includegraphics[width=0.95\linewidth,height=6cm]{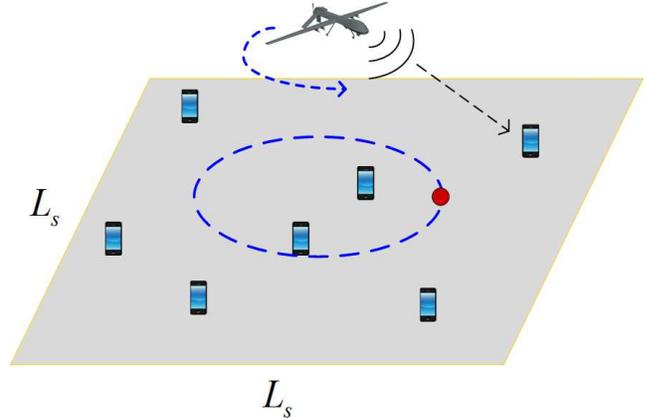}
\caption{Aerial BS serving delay-tolerant users}
\label{1}
\end{figure}
\section{System model and problem formulation}
\subsection{System Model}
As shown in Fig. 1, we consider a square geographical target area of dimension ${L_s}$ by ${L_s}$ containing a set of delay-tolerant ground users denoted by $\cal{M}$, where $|\cal{M}|$ = $M$, with $|.|$ denotes the cardinality. We assume the users have low-mobility and are uniformly distributed within the target area. The aerial BS is able to charge its battery at base, which is represented by the red dot as shown in Fig. 1. Therefore, within a given time period $T>0$, the aerial BS tries to cover as many ground users as possible before exhausting its on-board energy and flying back to the base. During any time period, the aerial BS serves its associated ground users via time-division multiple access (TDMA).  \\
\indent We consider a 3-D Cartesian coordinate system where the horizontal location of the $i$-th user in the set $\cal{M}$ is ${{\bf{w}}_i} = {[{x_i},{y_i}]^T} \in {\mathbb{R}^{2 \times 1}}$. We assume the aerial BS is flying with a fixed altitude $H$, where $H$ could correspond to the minimum altitude required for safe operation according to certain policies. For ease of exposition and following \cite{7888557,8247211}, we divide the total time period $T$ into $N$ equal time slots, indexed by $n = 1,2,...,N$. We further assume that the ground users can only be associated at these $N$ time slots. Note that the time slot length ${\delta _t}$ should be chosen to be efficiently small such that the location of the aerial BS changes only slightly within each time slot. Consequently, the trajectory, velocity and acceleration of the UAV are approximated by $N$ two-dimensional sequences as follows
\begin{eqnarray}
    &&{\bf{s}}\left[ n \right] \buildrel \Delta \over = {\bf{s}}(n{\delta _t}) = {[{s_x}[n],{s_y}[n]]^T},\\
    &&{\bf{v}}\left[ n \right] \buildrel \Delta \over = {\bf{v}}(n{\delta _t}) = {[{v_x}[n],{v_y}[n]]^T},\\
    &&{\bf{a}}\left[ n \right] \buildrel \Delta \over = {\bf{a}}(n{\delta _t}) = {[{a_x}[n],{a_y}[n]]^T},\\
    && n = 1,2,...,N\nonumber
\end{eqnarray}
In addition, the relationship among ${\bf{s}}\left[ n \right]$, ${\bf{v}}\left[ n \right]$ and ${\bf{a}}\left[ n \right]$ can be described by the following two equations \cite{7888557}
\begin{eqnarray}
    &&{\bf{v}}[n + 1] = {\bf{v}}[n] + {\bf{a}}[n]{\delta _t},\\
    &&{\bf{s}}[n + 1] = {\bf{s}}[n] + {\bf{v}}[n]{\delta _t} + \frac{1}{2}{\bf{a}}[n]{\delta _t}^2,\\
    && n = 1,2,...,N-1 \nonumber
\end{eqnarray}

For simplicity, we assume the air-to-ground (AtG) links are dominated by LoS channels \cite{7762053,7572068}. Note that the dominated LoS channels have not only been verified by field experiments\cite{qual}, but is also one of the main reasons that motivates us to deploy flying BS. Therefore, we have negligible small-scale effects and the channel quality is dominated by the communication distance. The distance from the aerial BS to user $i$ at time slot $n$ is given by
\begin{equation}
    {d_i}[n] = \sqrt {{H^2} + {{\left\| {{\bf{s}}[n] - {{\bf{w}}_i}} \right\|}^2}} 
\end{equation}
Correspondingly, the time-varying channel for user $i$ at the time slot $n$ is expressed as
\begin{equation}
    {h_i}[n] = \frac{{{\beta _0}}}{{{d_i}{{[n]}^2}}} = \frac{{{\beta _0}}}{{{H^2} + {{\left\| {{\bf{s}}[n] - {{\bf{w}}_i}} \right\|}^2}}}
\end{equation}
where we denote by ${{\beta _0}}$ the channel power at the reference distance ${d_0} = 1$ m. We define a binary variable ${\alpha _i}[n]$ indicating the scheduling and association status of user $i$ in time slot $n$. Specifically, the $i$-th user is served by the aerial BS at time slot $n$ if ${\alpha _i}[n] = 1$, and otherwise ${\alpha _i}[n] = 0$. We assume at most one of the $M$ users is associated with the aerial BS at each time slot, which can be expressed as
\begin{equation}
    \sum\limits_{i = 1}^M {{\alpha _i}[n]}  \le 1,\forall n
\end{equation}
Therefore, if user $i$ is scheduled for communicating with the aerial BS at time slot $n$, the signal-to-noise ratio (SNR) at user $i$ can be expressed as
\begin{equation}
    {\gamma _i}[n] = \frac{{P \cdot {h_i}[n]}}{{{\sigma ^2}}} = \frac{{P{\zeta _0}}}{{{H^2} + {{\left\| {{\bf{s}}[n] - {{\bf{w}}_i}} \right\|}^2}}}
\end{equation}
where $P$, ${{\sigma ^2}}$ and ${\zeta _0} = \frac{{{\beta _0}}}{{{\sigma ^2}}}$ denote the transmit power of the aerial BS, noise power and the referenced received signal-to-noise (SNR) respectively. The achievable total data for user $i$ in the unit of bits is thus given by
\begin{equation}
    {R_i} = \sum\limits_{n = 1}^N {{\alpha _i}[n]B{{\log }_2}} (1 + {\gamma _i}[n])
\end{equation}
where $B$ is the total available bandwidth. The total power consumption of the aerial BS consists of two parts, i.e., the power consumed for communication functions and the power consumed for supporting the mobility of UAV. In practice, the propulsion power consumption is much higher than the communication-related power, and we thus ignore the communication-related power consumption for simplicity \cite{7888557,8316986}. Propulsion power consumption depends on the flying status of UAV, and a theoretical model was derived in \cite{7888557}. For tractable analysis, the upper bound of the model is adopted in this paper, and the total consumed propulsion power can be expressed as 
\begin{equation}
    {P_{\rm{c}}} = \sum\limits_{n = 1}^N {({c_1}{{\left\| {{\bf{v}}[n]} \right\|}^3} + \frac{{{c_2}}}{{\left\| {{\bf{v}}[n]} \right\|}}(1 + \frac{{{{\left\| {{\bf{a}}[n]} \right\|}^2}}}{{{g^2}}}))} 
\end{equation}
where $c_1$ and $c_2$ are constant parameters related to the UAV's design, air density, etc., and $g = 9.8{\rm{ m/}}{{\rm{s}}^2}$ represents the gravitational acceleration. Correspondingly, the total consumed energy is expressed as
\begin{equation}
    {E_{\rm{c}}} = \sum\limits_{n = 1}^N {({c_1}{{\left\| {{\bf{v}}[n]} \right\|}^3} + \frac{{{c_2}}}{{\left\| {{\bf{v}}[n]} \right\|}}(1 + \frac{{{{\left\| {{\bf{a}}[n]} \right\|}^2}}}{{{g^2}}}))}  \cdot {\delta _t}
\end{equation}

\subsection{Problem Formulation}
Our goal is to maximize the number of covered ground users with a limited on-board energy by jointly optimizing the UAV trajectory and the user communication scheduling and association. We further define a binary variable ${\rho _i}$ indicating whether the data demand of user $i$ is satisfied or not. To be specific, if we denote the data requested by user $i$ as ${Q_i}$, ${\rho _i} = 1$ when ${R_i} \ge {Q_i}$, and otherwise ${\rho _i} = 0$. The optimization problem is then formulated as
\begin{subeqnarray}
    {\rm{(P1)}}:&&\mathop {{\rm{Maximize}}}\limits_{\left\{ {{\alpha _i}[n],{\bf{s}}[n],{\bf{v}}[n],{\bf{a}}[n],{\rho _i}} \right\}} \sum\limits_{i \in M} {{\rho _i}}  \\
    &&{\rm{subject}}{\kern 1pt}{\kern 1pt} {\rm{to}}\nonumber\\
    &&\sum\limits_{n = 1}^N {{\alpha _i}[n]B{{\log }_2}} (1 + {\gamma _i}[n]) \ge {\rho _i}{Q_i},\forall i\\
    &&{\rho _i} \in \left\{ {0,1} \right\},\forall i\\
    &&{\alpha _i}[n] \in \left\{ {0,1} \right\},\forall n,\forall i\\
    &&\sum\limits_{i = 1}^M {{\alpha _i}[n]}  \le 1,\forall n\\
    &&\sum\limits_{n = 1}^N {({c_1}{{\left\| {{\bf{v}}[n]} \right\|}^3} + \frac{{{c_2}}}{{\left\| {{\bf{v}}[n]} \right\|}}(1 + \frac{{{{\left\| {{\bf{a}}[n]} \right\|}^2}}}{{{g^2}}}))}  \cdot {\delta _t} \le {E_{{\rm{tot}}}}\nonumber\\
    &\\
    &&{\bf{s}}[n + 1] = {\bf{s}}[n] + {\bf{v}}[n]{\delta _t} + \frac{1}{2}{\bf{a}}[n]{\delta _t}^2, \nonumber\\
    &&n = 1,2,...,N - 1\\
    &&{\bf{v}}[n + 1] = {\bf{v}}[n] + {\bf{a}}[n]{\delta _t}, \nonumber\\
    &&n = 1,2,...,N - 1\\
    &&{\bf{s}}[0] = {\bf{s}}[N] = {{\bf{s}}_0}\\
    &&{\bf{v}}[0] = {{\bf{v}}_0}\\
    &&\left\| {{\bf{v}}[n]} \right\| \le {v_{\max }},\forall n\\
    &&\left\| {{\bf{v}}[n]} \right\| \ge {v_{\min }},\forall n\\
    &&\left\| {{\bf{a}}[n]} \right\| \le {a_{\max }},\forall n
\end{subeqnarray}
where ${E_{{\rm{tot}}}}$ denotes the total on-board energy of the aerial BS, ${{\bf{s}}_0}$ denotes the location of the base, ${{\bf{v}}_0}$, ${v_{\max }}$, ${v_{\min }}$ and ${a_{\max }}$ denote initial velocity, maximum allowed speed, minimum required speed and maximum allowed acceleration of the fixed-wing UAV respectively. As can be seen in the constraint (13b), when the achievable total data for user $i$ is equal or larger than the required data ${Q_i}$, ${\rho _i} = 1$ and the objective function is increased by one correspondingly. However, when the demand of user $i$ is not met, ${\rho _i} = 0$ and the objective function remains the same. Note that, (13f) guarantees that the total consumed energy should be no larger than the on-board energy of the UAV. According to (13i), the aerial BS is dispatched from the base at the first time slot, and should fly back to the base for recharging at the end of the mission period. In addition, the UAV mobility is governed by the velocity constraints as specified in (13k)-(13m). Notably, a minimum speed constraint is set for the aerial BS since it is impossible for fixed-wing UAVs to hover with zero speed. \\
\indent Problem P1 is a non-convex MINLP and is thus challenging to solve. Although the binary variables can be addressed with advanced mixed integer programming techniques, using solvers such as Gurobi and MOSEK \cite{7918510,8587183}, constraints (13b), (13f) and (13l) are non-convex and can not be straightforwardly solved.
 
\section{proposed iterative algorithm for coverage maximization}
In this section, we propose an efficient iterative algorithm based on block coordinate descent and successive convex optimization techniques to obtain the sub-optimal solution of P1. Define ${\bf{A}} = \left\{ {{\alpha _i}[n],\forall i,\forall n} \right\}$ and ${\bf{Q}} = \left\{ {{\bf{s}}[n],{\bf{v}}[n],{\bf{a}}[n],\forall n} \right\}$ as the user scheduling set and the UAV trajectory set respectively. For solving P1, we decompose the problem into two sub-problems and alternately optimize the two sub-problems within each iteration. Specifically, with a given UAV trajectory set ${\bf{Q}}$, first sub-problem of P1, which is denoted as P1.1 can be reformulated as
\begin{subeqnarray}
    {\rm{(P1}}{\rm{.1)}}:&&\mathop {{\rm{Maximize}}}\limits_{\left\{ {{\bf{A}},{\rho _i}} \right\}} \sum\limits_{i \in M} {{\rho _i}}\\
    &&{\rm{subject}}{\kern 1pt}{\kern 1pt} {\rm{to}}\nonumber\\
    &&\sum\limits_{n = 1}^N {{\alpha _i}[n]B{{\log }_2}} (1 + {\gamma _i}[n]) \ge {\rho _i}{Q_i},\forall i\\
    &&{\rho _i} \in \left\{ {0,1} \right\},\forall i\\
    &&{\alpha _i}[n] \in \left\{ {0,1} \right\},\forall n,\forall i\\
    &&\sum\limits_{i = 1}^M {{\alpha _i}[n]}  \le 1,\forall n
\end{subeqnarray}
Note that except the two constraints, i.e., (14c) and (14d), defining the boolean variables, (14a) is a linear objective function, and (14b) and (14e) are both linear constraints. Therefore, P1.1 is a mixed-integer linear problem (MILP), which can be solved efficiently by standard optimization solvers such as Gurobi and MOSEK.\\ 
\indent Similarly, by fixing the user scheduling variables ${\bf{A}}$, the UAV trajectory related variables ${\bf{Q}}$ can be optimized by solving the following sub-problem P1.2.
\begin{subeqnarray}
    {\rm{(P1}}{\rm{.2)}}:&&\mathop {{\rm{Maximize}}}\limits_{\left\{ {{\bf{Q}},{\rho _i}} \right\}} \sum\limits_{i \in M} {{\rho _i}} \\
    &&{\rm{subject}}{\kern 1pt}{\kern 1pt} {\rm{to}}\nonumber\\
    &&\sum\limits_{n = 1}^N {{\alpha _i}[n]B{{\log }_2}} (1 + {\gamma _i}[n]) \ge {\rho _i}{Q_i},\forall i\\
    &&\sum\limits_{n = 1}^N {({c_1}{{\left\| {{\bf{v}}[n]} \right\|}^3} + \frac{{{c_2}}}{{\left\| {{\bf{v}}[n]} \right\|}}(1 + \frac{{{{\left\| {{\bf{a}}[n]} \right\|}^2}}}{{{g^2}}}))}  \cdot {\delta _t} \le {E_{{\rm{tot}}}}\nonumber\\
    &\\
    &&{\bf{s}}[n + 1] = {\bf{s}}[n] + {\bf{v}}[n]{\delta _t} + \frac{1}{2}{\bf{a}}[n]{\delta _t}^2, \nonumber\\
    &&n = 1,2,...,N - 1\\
    &&{\bf{v}}[n + 1] = {\bf{v}}[n] + {\bf{a}}[n]{\delta _t}, \nonumber\\
    &&n = 1,2,...,N - 1\\
    &&{\bf{s}}[0] = {\bf{s}}[N] = {{\bf{s}}_0}\\
    &&{\bf{v}}[0] = {{\bf{v}}_0}\\
    &&\left\| {{\bf{v}}[n]} \right\| \le {v_{\max }},\forall n\\
    &&\left\| {{\bf{v}}[n]} \right\| \ge {v_{\min }},\forall n\\
    &&\left\| {{\bf{a}}[n]} \right\| \le {a_{\max }},\forall n\\
    &&{\rho _i} \in \left\{ {0,1} \right\},\forall i
\end{subeqnarray}
Note that constraints (15d)-(15g) are linear, (15h) and (15j) are convex and (15k) specifies that ${\rho _i}$ is a boolean variable, so the difficulty of solving P1.2 lies in constraints (15b), (15c) and (15i), which are all non-convex. We first observe that, although the left-hand-side (LHS) of constraint (15b), which is ${R_i}$, is not concave with respect to ${\bf{s}}[n]$, it is convex with respect to ${{{\left\| {{\bf{s}}[n] - {{\bf{w}}_i}} \right\|}^2}}$. Since any convex function is globally lower-bounded by its first order Taylor expansion at any point \cite{Boyd:2004:CO:993483}, successive convex optimization technique can be applied to tackle (15b). To be specific, with a given local UAV location $\{ {{\bf{s}}_l}[n],\forall n\}$, we yield the following lower bound $R_i^{{\rm{lb}}}$ for ${R_i}$
\begin{eqnarray}
    &&{R_i} = \sum\limits_{n = 1}^N {{\alpha _i}[n]B{{\log }_2}} (1 + \frac{{P{\zeta _0}}}{{{H^2} + {{\left\| {{\bf{s}}[n] - {{\bf{w}}_i}} \right\|}^2}}})\nonumber\\
    &&   \ge  - \sum\limits_{n = 1}^N {{\alpha _i}[n]B \cdot } A_i^l[n]\left( {{{\left\| {{\bf{s}}[n] - {{\bf{w}}_i}} \right\|}^2} - {{\left\| {{{\bf{s}}_l}[n] - {{\bf{w}}_i}} \right\|}^2}} \right)\nonumber\\
    &&  + \sum\limits_{n = 1}^N {{\alpha _i}[n]B \cdot } B_i^l[n] \buildrel \Delta \over = R_i^{{\rm{lb}}}
\end{eqnarray}
where $A_i^l[n]$ and $B_i^l[n]$ are constants which are given by
\begin{equation}
    A_i^l[n] = \frac{{({{\log }_2}e)P{\zeta _0}}}{{({H^2} + {{\left\| {{{\bf{s}}_l}[n] - {{\bf{w}}_i}} \right\|}^2})({H^2} + {{\left\| {{{\bf{s}}_l}[n] - {{\bf{w}}_i}} \right\|}^2} + P{\zeta _0})}} 
\end{equation}
\begin{equation}
    B_i^l[n] = {\log _2}(1 + \frac{{P{\zeta _0}}}{{{H^2} + {{\left\| {{{\bf{s}}_l}[n] - {{\bf{w}}_i}} \right\|}^2}}}),\forall n,\forall i
\end{equation}
The equality of (16) holds at the point ${\bf{s}}[n] = {{\bf{s}}_l}[n],\forall n$. With the use of $R_i^{{\rm{lb}}}$, the non-convex constraint (15b) can be reformulated as 
\begin{equation}
    R_i^{{\rm{lb}}} \ge {\rho _i}{Q_i},\forall i
\end{equation}
Since $R_i^{{\rm{lb}}}$ is a concave function with respect to ${{\bf{s}}[n]}$, (19) is convex now. Furthermore, for addressing the non-convexity of (15c) and (15i), we introduce slack variables $\{ {\tau _n}\}$ as in \cite{7888557,8329973}, and the corresponding new constraints are
\begin{eqnarray}
    &&\sum\limits_{n = 1}^N {({c_1}{{\left\| {{\bf{v}}[n]} \right\|}^3} + \frac{{{c_2}}}{{{\tau _n}}}(1 + \frac{{{{\left\| {{\bf{a}}[n]} \right\|}^2}}}{{{g^2}}}))} \cdot {\delta _t} \le {E_{{\rm{tot}}}}\nonumber\\
    &\\
    &&{\tau _n} \ge {v_{\min }},\forall n\\
    &&{\left\| {{\bf{v}}[n]} \right\|^2} \ge {\tau _n}^2,\forall n
\end{eqnarray}
With the introduced slack variables $\{ {\tau _n}\}$, variable ${\bf{v}}\left[ n \right]$ and ${\bf{a}}\left[ n \right]$ are no more coupled, and the LHS of constraint (20) is now jointly convex with respect to $\{ {\bf{v}}\left[ n \right],{\bf{a}}\left[ n \right],{\tau _n}\}$. Note that with such a relaxation, a new non-convex constraint (22) is introduced. Fortunately, a local optimal solution can be obtained by applying successive convex optimization. Specifically, since the LHS of (22) is convex and differentiable with respect to ${\bf{v}}\left[ n \right]$, a lower-bound of ${\left\| {{\bf{v}}[n]} \right\|^2}$ can be obtained with any given local point $\{ {{\bf{v}}_l}[n],\forall n\} $ by leveraging the first-order Taylor expansion of ${\left\| {{\bf{v}}[n]} \right\|^2}$ as follows
\begin{eqnarray}
&&{\left\| {{\bf{v}}[n]} \right\|^2} \ge {\left\| {{{\bf{v}}_l}[n]} \right\|^2} + 2{\bf{v}}_l^T[n]\left( {{\bf{v}}[n] - {{\bf{v}}_l}[n]} \right)\nonumber\\
&& \buildrel \Delta \over = {\psi _{{\rm{lb}}}}({\bf{v}}[n])
\end{eqnarray}
where the equality holds at the point ${\bf{v}}[n] = {{\bf{v}}_l}[n],\forall n$. Therefore, we replace (22) with the following new convex constraint 
\begin{equation}
    {\psi _{{\rm{lb}}}}({\bf{v}}[n]) \ge {\tau _n}^2,\forall n
\end{equation}
The sub-problem P1.2 can thus be reformulated as
\begin{subeqnarray}
    {\rm{(P}}{1.2'}):&&\mathop {{\rm{Maximize}}}\limits_{\left\{ {{\bf{Q}},{\rho _i},{\tau _n}} \right\}} \sum\limits_{i \in M} {{\rho _i}}  \\
    &&{\rm{subject}}{\kern 1pt}{\kern 1pt} {\rm{to}} {\kern 1pt}{\kern 1pt}{\kern 1pt}{\kern 1pt}(15{\rm{d}}-15{\rm{h}}),(15{\rm{j}}),(15{\rm{k}})\nonumber\\
    &&R_i^{{\rm{lb}}} \ge {\rho _i}{Q_i},\forall i\\
    &&\sum\limits_{n = 1}^N {({c_1}{{\left\| {{\bf{v}}[n]} \right\|}^3} + \frac{{{c_2}}}{{{\tau _n}}}(1 + \frac{{{{\left\| {{\bf{a}}[n]} \right\|}^2}}}{{{g^2}}}))} \cdot {\delta _t} \le {E_{{\rm{tot}}}}\nonumber\\
    &\\
    &&{\tau _n} \ge {v_{\min }},\forall n\\
    &&{\psi _{{\rm{lb}}}}({\bf{v}}[n]) \ge {\tau _n}^2,\forall n
\end{subeqnarray}
Since all constraints of ${\rm{P}}{1.2'}$ are convex and the objective function is a MILP, the optimization problem can again be efficiently solved by standard optimization solvers.\\
\indent Based on the solution of the two sub-problems ${\rm{P1.1}}$ and ${\rm{P1.2'}}$, we propose an iterative algorithm by applying block coordinate descent method for solving ${\rm{P1}}$. To be specific, the optimization variables of the original problem are partitioned into two blocks ${\bf{A}}$ and ${\bf{Q}}$ as defined in the beginning of this part. The user scheduling and association variable ${\bf{A}}$ and the user trajectory related variables ${\bf{Q}}$ are then alternately optimized by solving ${\rm{P1}}{\rm{.1}}$ and ${\rm{P1.2'}}$ correspondingly, while keeping the other block of variables fixed. Additionally, the optimized variables in each iteration are served as inputs of the next iteration until there is no increase in objective value any more after a certain iteration. For simplicity, the whole iterative algorithm is summarized in Algorithm 1.\\
\begin{algorithm}[t!]
\caption{Block coordinate descent technique for solving P1}
\begin{algorithmic}[1]
\renewcommand{\algorithmicrequire}{\textbf{Initialization:}} 
\REQUIRE
Initial the trajectory set ${{\bf{Q}}_{\rm{0}}}$; Let $l = 0$; 
\REPEAT
\STATE
solve problem ${\rm{P1}}{\rm{.1}}$ with given $\{ {{\bf{Q}}_l}\} $, and denote the optimal solution as $\{ {{\bf{A}}_{l + 1}}\}$
\STATE
solve problem ${\rm{P1.2'}}$ with given $\{ {{\bf{A}}_{l + 1}}\}$, and denote the optimal solution as $\{ {{\bf{Q}}_{l + 1}}\}$ 
\STATE
update $l = l + 1$.
\UNTIL the objective value keeps the same as the value obtained in the previous iteration
\end{algorithmic}
\end{algorithm}
\indent In the following, we prove the convergence of Algorithm 1. Define $\eta ({{\bf{A}}_l},{{\bf{Q}}_l})$ and $\eta _{{\rm{trj}}}^{{\rm{lb}}}({{\bf{ A}}_l},{{\bf{Q}}_l})$ as the objective value of ${\rm{P1}}$ and ${\rm{P1.2'}}$ respectively. It then follows that
\begin{eqnarray}
    \eta ({{\bf{{\rm A}}}_l},{{\bf{Q}}_l})&\mathop  \le \limits^a& \eta ({{\bf{{\rm A}}}_{l + 1}},{{\bf{Q}}_l})\nonumber\\
    &\mathop  = \limits^b& \eta _{{\rm{trj}}}^{{\rm{lb}}}({{\bf{{\rm A}}}_{l + 1}},{{\bf{Q}}_l})\nonumber\\
    &\mathop  \le \limits^c& \eta _{{\rm{trj}}}^{{\rm{lb}}}({{\bf{{\rm A}}}_{l + 1}},{{\bf{Q}}_{l + 1}})\nonumber\\
    &\mathop  \le \limits^d& \eta ({{\bf{{\rm A}}}_{l + 1}},{{\bf{Q}}_{l + 1}})
\end{eqnarray}
where (a) holds since in step 2 of Algorithm 1, the optimal solution of ${\rm{P1}}{\rm{.1}}$ , which is ${{\bf{A}}_{l + 1}}$, is obtained based on given ${{\bf{Q}}_l}$; (b) holds due to the fact that the first order Taylor expansions in (16) and (23) are tight at the given local location and the given local velocity respectively, so ${\rm{P1}}{\rm{.2}}$ and ${\rm{P1.2'}}$ has the identical objective value; (c) holds since with the given ${{\bf{A}}_{l + 1}}$ and ${{\bf{Q}}_l}$, ${\rm{P1.2'}}$ is optimally solved in step 3 of Algorithm 1 with solution ${{\bf{Q}}_{l + 1}}$; (d) holds as the objective value obtained by solving ${\rm{P1.2'}}$ serves as the lower-bound of that of the original problem ${\rm{P1}}{\rm{.}}{{\rm{2}}}$ at ${{\bf{Q}}_{l + 1}}$. Therefore, (26) suggests that the proposed algorithm is non-decreasing. In addition, since the objective value of ${\rm{P1}}$ is clearly upper-bounded by a finite integer value, which corresponds to the total number of ground users, the algorithm is guaranteed to converge.

\section{Initial trajectory design: Circular vs Tailored Path}
The prerequisite for applying Algorithm 1 is initializing the trajectory set ${\bf{Q}}$. According to \cite{7366709,8443133}, both the converged solution and performance of such iterative algorithm depend on the initialization schemes. Therefore, for achieving faster convergence speed and better user coverage performance, we design a simple initial trajectory for Algorithm 1 in this section. \\
\indent Since the dispatched aerial BS has to return to the base for recharging within a given time period, the typical initial trajectory for such scenario is a circular trajectory \cite{8247211,8329013} which serves as the benchmark. Specifically, for the ${L_s}$ by ${L_s}$ square geographical target area, we assume the center of the circular initial trajectory (CIT) is ${{\bf{c}}_{\rm{t}}} = {[\frac{{{L_s}}}{2},\frac{{{L_s}}}{2}]^T \in {\mathbb{R}^{2 \times 1}}}$ and the radius of the trajectory is ${r_{\rm{t}}} = \frac{{{L_s}}}{4}$ so the number of users inside and outside the trajectory is balanced. In addition, we assume the base is located at ${{\bf{s}}_{\rm{b}}} = {{\bf{c}}_{\rm{t}}} + {[{r_{\rm{t}}},0]^T} = {[\frac{{{L_s}}}{2} + {r_{\rm{t}}},\frac{{{L_s}}}{2}]^T}$. \\
\begin{figure}[t!]
\centering
\includegraphics[width=1\linewidth]{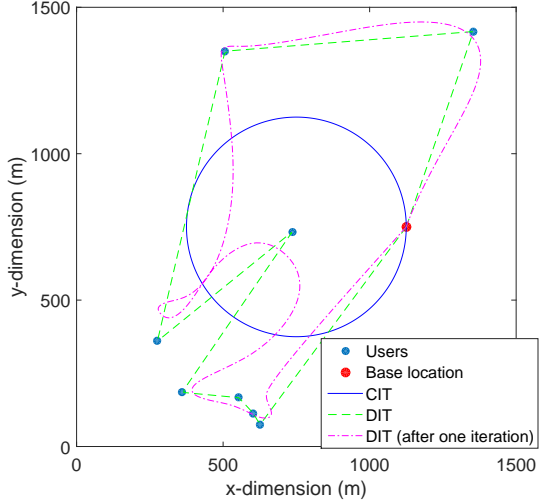}
\caption{An example of CIT, DIT and the generated trajectory after one iteration of Algorithm 1 with DIT, $T = 100$s, ${E_{{\rm{tot}}}} = 1.5 \times {10^4}$J}
\label{2}
\end{figure}
\indent Different from most of the UAV trajectory design problems, where the aerial BS associates with all the ground users, e.g. \cite{8247211}, only part of the ground users can be scheduled and associated in our specific problem. In this case, if CIT is applied to Algorithm 1, users which are closer to the initial trajectory has a higher opportunity to be considered for association due to the lower path loss. In addition, users which are not scheduled in the first iteration will only be considered for association when the the demand of all the scheduled users are satisfied after optimizing the trajectory. Therefore, CIT does not consider fair scheduling and association and may lead to a performance loss. This motivates us to design an initial trajectory which ensures all the ground users can get close to the UAV in certain time slots, so the users have a relatively fair opportunity to be considered for scheduling and association. To this aim, we design an initial trajectory where the UAV flies straightly from one ground user to the other with constant speed $\left\| {{\bf{v}}[n]} \right\| = V$ in the horizontal dimension, and finally backs to the base. Specifically, the designed initial trajectory (DIT) is summarized as follows 
\begin{enumerate}
    \item  Convert the location of ground users into polar coordinate system with ${{\bf{c}}_{\rm{t}}}$ serves as the coordinate origin, that is ${\bf{w}}_i^{\rm{p}} = {[{r_i},{\theta _i}]^T}$, where ${r_i} = \left\| {{{\bf{c}}_{\rm{t}}} - {{\bf{w}}_i}} \right\|$ and ${\theta _i} = \arctan (\frac{{{y_i} - \frac{{{L_s}}}{2}}}{{{x_i} - \frac{{{L_s}}}{2}}}) \in \left( {{\rm{0, 2}}\pi } \right)$.
    \item Starting from the charging base location, the initialization path connects each of the ground users with a straight line based on a counterclockwise order. If two users have the same ${\theta _i}$, the initialization path prioritizes the user which has a smaller $r_i$.
    \item Resort all the $M$ users according to the access order in step 2, such that the first ground user is the one which has the smallest ${\theta _i}$ .
    \item Calculate the total distance of DIT, which is
    \begin{equation}
        {d_{{\rm{sum}}}} = \sum\limits_{i = 1}^{M - 1} {\left\| {{{\bf{w}}_{i + 1}} - {{\bf{w}}_i}} \right\|}  + \left\| {{{\bf{s}}_{\rm{b}}} - {{\bf{w}}_1}} \right\| + \left\| {{{\bf{s}}_{\rm{b}}} - {{\bf{w}}_M}} \right\|
    \end{equation}
    \item The distance interval is then calculated as ${\delta _d} = \frac{{{d_{{\rm{sum}}}}}}{N}$, and the initial trajectory can be obtained accordingly.
\end{enumerate}
Intuitively, the third step of Algorithm 1 forces the UAV to fly closer to the scheduled ground users in the corresponding time slots, so more requested data can be downloaded thanks to the decreased path loss. Since the UAV has a much closer distance with the associated ground users by applying DIT, the proposed initial trajectory is also expected to speed up the convergence. However, note that DIT does not necessarily satisfy the UAV energy constraint (13f) and the mobility constraints (13k-13m). Fortunately, the third step of Algorithm 1 guarantees to generate a trajectory which satisfies all the constraints shown in ${\rm{P1.2'}}$, and the generated trajectory is based on a much fairer scheduling and association scheme compared to CIT. Therefore, the performance of Algorithm 1 is still non-decreasing and thus converges from the second iteration. For better illustration of the proposed initial trajectory, Fig. 2 compares CIT, DIT and the generated trajectory after one iteration of Algorithm 1 by applying DIT. Note that, the users which are located far away from CIT ,e.g., the one in the top right corner may never be scheduled and associated by applying Algorithm 1 with CIT due to the large path loss. On the contrary, these users could be served by applying Algorithm 1 with DIT thanks to the significantly reduced transmission distance.

\section{Imperfect ULI and robust optimization}
In real scenarios, the accuracy of GPS systems is affected by lots of factors such as weather and terrain \cite{article}. Consequently, the number of served users may decrease drastically in the existence of inaccurate ULI. In this section, we propose two robust techniques for compensating the performance loss when user location is estimated inaccurately.   

\subsection{Worst Case (WC) ULI Robust Optimization}
We first model the estimated user location as $\widetilde {{{\bf{w}}_i}} = {[{x_i} + {e_{xi}},{y_i} + {e_{yi}}]^T}$, where ${e_{xi}}$ and ${e_{yi}}$ denote the estimation error in the x-axis and y-axis respectively. Both ${e_{xi}}$ and ${e_{yi}}$ follow Gaussian distribution with zero mean and standard deviation $\sigma $ in meters. We assume that the maximum deviation between real user location and the estimated user location is ${d_{{\rm{th}}}}$, where ${d_{{\rm{th}}}} \approx 3\sigma$. Therefore, the real position of user $i$ is bounded by a circle region with radius ${d_{{\rm{th}}}}$ and circle center $\widetilde {{{\bf{w}}_i}}$. 
\begin{figure}[t!]
\centering
\includegraphics[width=0.75\linewidth, height=6cm]{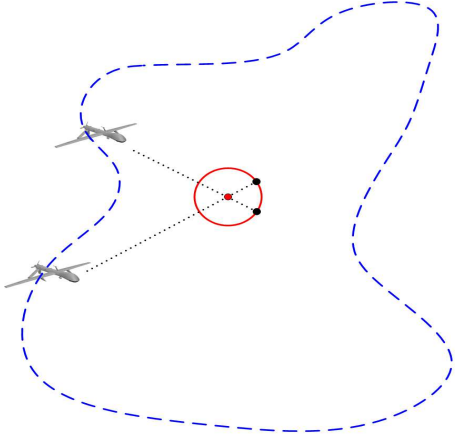}
\caption{Optimizing the trajectory with respect to the worst case ULI}
\label{3}
\end{figure}
For increasing the robustness against inaccurate ULI, we first propose a simple technique which guarantees the coverage performance in the worst case. Instead of solving ${\rm{P1}}$ with $\widetilde {{{\bf{w}}_i}}$, we employ the worst case ULI estimate. The worst case user location at a specific time slot is the farthest intersection between the circle which specifies the region of actual user location and a straight line which starts from the UAV position ${\bf{s}}\left[ n \right]$ and passes through $\widetilde {{{\bf{w}}_i}}$. As shown in Fig. 3, the red dot represents the estimated user location and the worst case user location which is represented by the black dot is the farthest intersection point between the line and the red circle. Correspondingly, instead of solving ${\rm{P1}}$, the proposed robust technique tries to find the optimal trajectory and optimal scheduling and association by solving the following problem
\begin{subeqnarray}
    {\rm{(P2)}}:&&\mathop {{\rm{Maximize}}}\limits_{\left\{ {{\alpha _i}[n],{\bf{s}}[n],{\bf{v}}[n],{\bf{a}}[n],{\rho _i}} \right\}} \sum\limits_{i \in M} {{\rho _i}}  \\
    &&{\rm{subject}}{\kern 1pt}{\kern 1pt} {\rm{to}}{\kern 1pt}{\kern 1pt}{\kern 1pt}{\kern 1pt} (13\rm{c}) - (13\rm{m}) \nonumber\\
    &&\sum\limits_{n = 1}^N {{\alpha _i}[n]B{{\log }_2}} (1 + \widetilde {{\gamma _i}}[n]) \ge {\rho _i}{Q_i},\forall i
\end{subeqnarray}
where we have
\begin{equation}
    \widetilde {{\gamma _i}}[n]{\rm{ = }}\frac{{P \cdot {\zeta _0}}}{{{H^2} + {{(\left\| {{\bf{s}}[n] - \widetilde {{{\bf{w}}_i}}} \right\| + {d_{{\rm{th}}}})}^2}}}
\end{equation}
Since the LHS of (28b) is convex with respect to ${{{(\left\| {{\bf{s}}[n] - \widetilde {{{\bf{w}}_i}}} \right\| + {d_{{\rm{th}}}})}^2}}$ when the association variable is given, ${\rm{P2}}$ can be solved by the same successive convex optimization and block coordinate descent techniques as we used for solving ${\rm{P1}}$. With the optimal solutions of ${\rm{P2}}$, all the users that are covered in ${\rm{P2}}$ are guaranteed to be covered in ${\rm{P1}}$ with inaccurate ULI since the former considers the worst case performance. 

\subsection{Robust Optimization based on Minimum Excess Data Maximization (MEDM)}
In the preceding subsection, robustness is increased by guaranteeing the worst case conditions. In this subsection, we increase the robustness against inaccurate ULI from another perspective. We first note that once Algorithm 1 gives an optimal solution such that ${M_{\rm{s}}}$ users are served, where ${M_{\rm{s}}} \le {M}$, ${N_{\rm{s}}}$ out of $N$ time slots are allocated for satisfying the requirement of the ${M_{\rm{s}}}$ users. In other words, instead of providing more bits to the covered users, the aerial BS tries to allocate redundant time slots to users which can not be fully served. We assume the set which contains all the covered users is denoted by ${\cal{S}}$. Therefore, for the $m$-th user in the set ${\cal{S}}$, we have ${R_m} \ge {Q_m}$. Since increased immunity to inaccurate ULI for the $m$-th covered user can be achieved by increasing the excessive serving data ${\epsilon _m} = {R_m} - {Q_m}$, we propose another robust technique which maximizes the minimum excessive data among covered users. The optimization problem is formulated as 
\begin{subeqnarray}
{\rm{(P3)}}:&&\mathop {{\rm{Max}}\min }\limits_{\{ {\alpha _m}[n],{\bf{s}}[n],{\bf{v}}[n],{\bf{a}}[n]\} } ({R_m} - {Q_m})\\
 &&{\rm{subject}}{\kern 1pt}{\kern 1pt} {\rm{to}}{\kern 1pt}{\kern 1pt}{\kern 1pt}{\kern 1pt} (13\rm{f})-(13\rm{m}) \nonumber\\
&&{\alpha _m}[n] \in \left\{ {0,1} \right\},\forall n,\forall m\\
&&\sum\limits_{m = 1}^{M_{\rm{s}}} {{\alpha _m}[n]}  \le 1,\forall n
\end{subeqnarray}
where 
\begin{equation}
    {R_m} = \sum\limits_{n = 1}^N {{\alpha _m}[n]B{{\log }_2}} (1 + {\gamma _m}[n])
\end{equation}
\begin{equation}
    {\gamma _m}[n] = \frac{{P \cdot {\zeta _0}}}{{{H^2} + {{\left\| {{\bf{s}}[n] - {{\bf{w}}_m}} \right\|}^2}}},m \in \cal{S}
\end{equation}
Note that all the time slots are allocated to the covered users in ${\rm{P}}3$, and the trajectory and association variables are optimized for increasing the minimum ${\epsilon_m}$. The above optimization problem is equivalent to maximizing the auxiliary variable $\eta$ representing the minimum excessive data according to
\begin{subeqnarray}
{\rm{(P3.1)}}:&&\mathop {\max }\limits_{\{ {\alpha _m}[n],{\bf{s}}[n],{\bf{v}}[n],{\bf{a}}[n],\eta \} } \eta  \\
 &&{\rm{subject}}{\kern 1pt}{\kern 1pt} {\rm{to}}{\kern 1pt}{\kern 1pt}{\kern 1pt}{\kern 1pt} (13\rm{f})-(13\rm{m}) \nonumber\\
 &&{R_m} - {Q_m} \ge \eta ,\forall m\\
&&{\alpha _m}[n] \in \left\{ {0,1} \right\},\forall n,\forall m\\
&&\sum\limits_{m = 1}^{M_{\rm{s}}} {{\alpha _m}[n]}  \le 1,\forall n
\end{subeqnarray}
\begin{table}[t!]
\centering
\caption{Simulation parameters}
\label{my-label}
\begin{tabular}{|l|l|l|l|}
\hline 
 parameter&value&parameter&value\\
 \hline 
 $B$&${10^6}$ Hz&$H$&100 m  \\
 \hline 
 P&0.01 W&$g$&9.8 ${\rm{m}}/{{\rm{s}}^2}$   \\
 \hline 
 ${\beta _0}$&-50 dB&${\delta _t}$ & 0.5 s  \\
 \hline
 ${\sigma ^2}$&-110 dBm&${v_{\max }}$&80 m/s\\
 \hline
 ${c_1}$&$9.26 \times {10^{ - 4}}$&${v_{\min }}$&3 m/s \\
 \hline
 ${c_2}$&2250&${a_{\max }}$&6 ${\rm{m}}/{{\rm{s}}^2}$ \\
 \hline
\end{tabular}
\end{table}
Note that the non-convex constraint (33b) can be tackled with the same method as shown in (16)-(18), which yields
\begin{equation}
    R_m^{{\rm{lb}}} - {Q_m} \ge \eta ,\forall m
\end{equation}
Here, $R_m^{{\rm{lb}}}$ is the first order Taylor approximation of $R_m$ and denotes its lower bound. 
\begin{eqnarray}
    && R_m^{{\rm{lb}}}   \buildrel \Delta \over =  \sum\limits_{n = 1}^N {{\alpha _m}[n]B \cdot } B_m^l[n] \nonumber\\
    &&  -  \sum\limits_{n = 1}^N {{\alpha _m}[n]B \cdot } A_m^l[n]\left( {{{\left\| {{\bf{s}}[n] - {{\bf{w}}_m}} \right\|}^2} - {{\left\| {{{\bf{s}}_l}[n] - {{\bf{w}}_m}} \right\|}^2}} \right)\nonumber\\
\end{eqnarray}
where $A_m^l[n]$ and $B_m^l[n]$ are constants which are given by
\begin{equation}
    A_m^l[n] = \frac{{({{\log }_2}e)P{\zeta _0}}}{{({H^2} + {{\left\| {{{\bf{s}}_l}[n] - {{\bf{w}}_m}} \right\|}^2})({H^2} + {{\left\| {{{\bf{s}}_l}[n] - {{\bf{w}}_m}} \right\|}^2} + P{\zeta _0})}} 
\end{equation}
\begin{equation}
    B_m^l[n] = {\log _2}(1 + \frac{{P{\zeta _0}}}{{{H^2} + {{\left\| {{{\bf{s}}_l}[n] - {{\bf{w}}_m}} \right\|}^2}}}),\forall n,\forall m
\end{equation}
\begin{figure}[t!]
\centering
\includegraphics[width=0.85\linewidth, height=6cm]{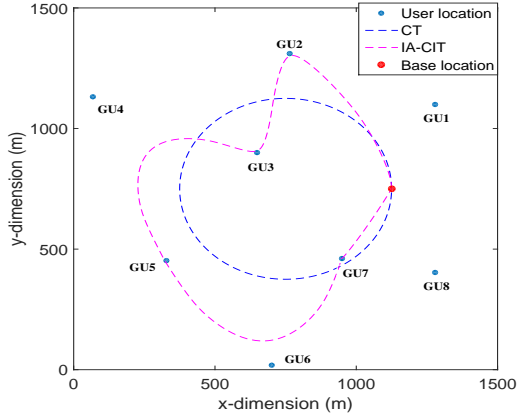}
\caption{Optimized trajectory with IA-CIT, $T$=100 s, ${E_{{\rm{tot}}}} = 1.5 \times {10^4}$ J }
\label{4}
\end{figure}
\begin{figure}[t!]
\centering
\includegraphics[width=0.8\linewidth, height=6.2cm]{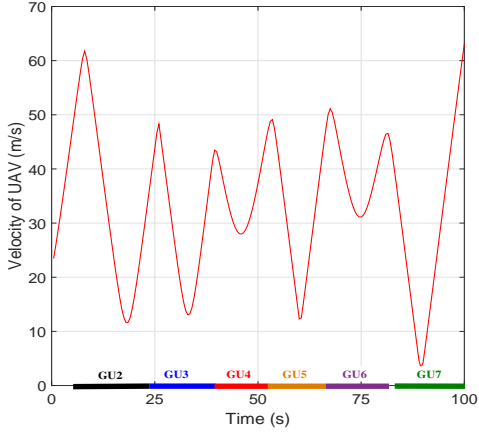}
\caption{Speed of aerial BS corresponding to the trajectory shown in Fig. 4}
\label{5}
\end{figure}
Except the objective function and constraint (33b), the only difference between problem ${\rm{P}}3.1$ and ${\rm{P}}1$ is that the time slots in ${\rm{P}}3.1$ can only be allocated to the severed users. Since the other two non-convex constraints in ${\rm{P}}3.1$, which are (13f) and (13l) have already been addressed in (20)-(24), we apply the same block coordinate descent and successive convex optimization techniques to solve ${\rm{P}}3.1$ as Algorithm 1. Note that for solving ${\rm{P}}3.1$, the iterative algorithm repeats until the fractional increase of the objective value is below a certain threshold $\varepsilon  > 0$. With this robust technique, all the covered users receive more bits than required demand. Therefore, even less bits are provided by the aerial BS at some time slots due to inaccurate ULI, the corresponding users are still covered as long as $\widetilde {{R_m}} \ge {Q_m},\forall m$, where $\widetilde {{R_m}}$ denotes the actual sum data provided to user $m$.

\section{simulation results and analysis}
In this section, we provide numerical results to evaluate the performance of our proposed techniques. We assume $M = 8$ users are distributed randomly within the square target area of $1.5 \times 1.5 {\rm{ k}}{{\rm{m}}^2}$. Correspondingly, the charging base is located in ${[1125,750]^T}$ and the radius of CIT is ${r_{\rm{t}}} = 375$m. Unless otherwise stated, we use the parameters shown in Table 1. In addition, the data demand of each user is a random value within the range of $[1,20]$ Mbits. The coverage performance is evaluated with regard to user coverage probability, which is defined as the ratio of number of served users to the total number of ground users within the target area. With a given number of ground users, increased coverage probability is clearly obtained by increasing the number of covered user. For ease of presentation, the proposed iterative algorithm with CIT and DIT are termed as IA-CIT and IA-DIT respectively. Furthermore, the proposed first robust technique which considers the worst case performance and the proposed second robust technique which maximizes the minimum excessive data are namely WC and MEDM correspondingly. 
\begin{figure}[t!]
\centering
\includegraphics[width=0.85\linewidth, height=6cm]{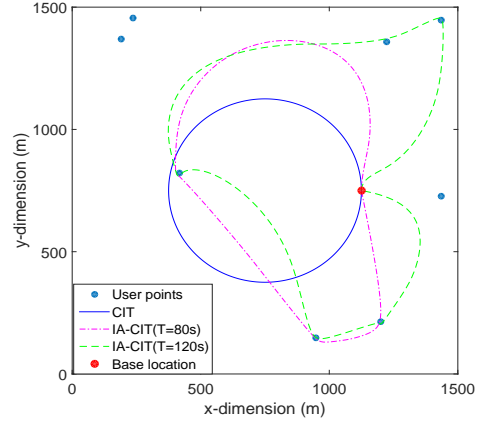}
\caption{Optimized trajectory with IA-CIT for different time period $T$, ${E_{{\rm{tot}}}} = 2.5 \times {10^4}$ J }
\label{6}
\end{figure}
\begin{figure}[t!]
\centering
\includegraphics[width=0.85\linewidth, height=5.75cm]{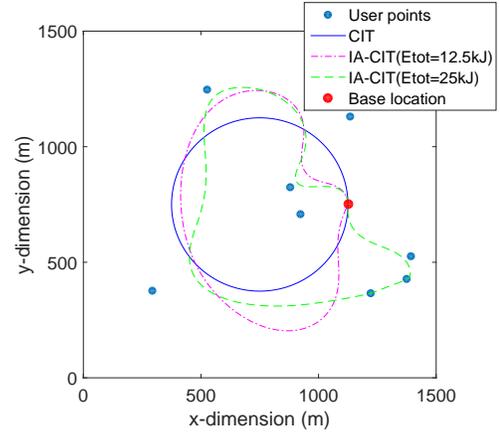}
\caption{Optimized trajectory with IA-CIT for different on-board energy ${E_{{\rm{tot}}}}$, $T$=120 s  }
\label{7}
\end{figure}
\subsection{Proposed Iterative Algorithm and the Impact of Time and Energy Constraints}
In Fig. 4, we first illustrate the optimized trajectory obtained by the proposed IA-CIT, assuming $T$=100 s and ${E_{{\rm{tot}}}} = 1.5 \times {10^4}$ J. It can be seen that for the ground users which are covered, the UAV tries to move close to them to reduce path loss and thus transmits more information. Although different users have various data request, the users which are located closer to the CIT have a better opportunity to be scheduled and associated thanks to the clearly better AtG communication channel. The UAV trajectory is smooth, and this is a result of UAV acceleration constraint, which forbids the UAV to change its direction abruptly. For better understanding the aerial BS's flying status, Fig. 5 shows the time-varying UAV speed as well as the user scheduling and association corresponding to the trajectory shown in Fig. 4. It can be seen that the UAV first flies towards the served users with increased speed, then gradually reduces its speed when it starts to have a good AtG communication channel with the corresponding users. Since fixed-wing UAV is utilized, the aerial BS has a minimum speed requirement for maintaining the movement and can not hover above the served users with zero speed. Note that not all the time slots are allocated for the covered users. This indicates that the aerial BS tries to allocate redundant time slots to users which can not be fully served after satisfying the requirement of the covered users.\\
\begin{figure}[t!]
\centering
\includegraphics[width=0.8\linewidth, height=6cm]{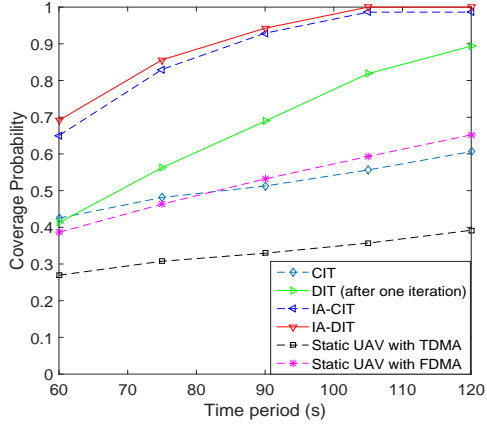}
\caption{Coverage probability versus time period $T$ with different techniques, ${E_{{\rm{tot}}}} = 2.5 \times {10^4}$J}
\label{8}
\end{figure}
\indent The number of covered users is restricted by the time period $T$ and the limited on-board energy resources ${E_{{\rm{tot}}}}$. Firstly, Fig. 6 illustrates the optimized trajectories obtained by IA-CIT under different $T$ with large enough on-board energy ${E_{{\rm{tot}}}} = 2.5 \times {10^4}$J. As shown in Fig. 6, with a larger time period, request of more ground users can be satisfied as more time slots are allocated for transmitting information.  Ideally, it is expected that all the ground users can be covered when $T$ is large enough. However, increasing $T$ not only increases the user access delay but also increases the consumed energy. As $T$ increases, each served user needs to wait for a longer time to be associated and more built-in energy is consumed. Therefore, in real scenarios, the choice of $T$ should take both time-delay tradeoff and energy consumption into account. \\
\begin{figure}[t!]
\centering
\includegraphics[width=0.8\linewidth, height=6cm]{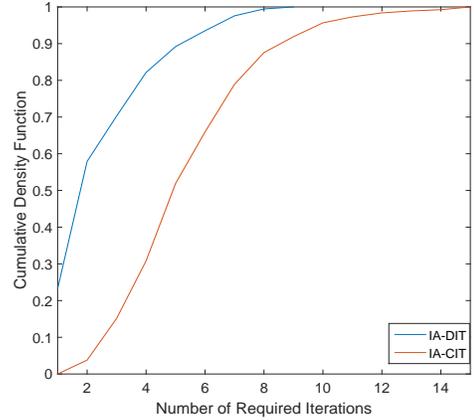}
\caption{CDF of number of required iterations for IA-CIT and IA-DIT, ${E_{{\rm{tot}}}} = 2.5 \times {10^4}$J, $T$=120 s  }
\label{9}
\end{figure}
\indent On the contrary, Fig. 7 shows the optimized trajectories obtained by IA-CIT with various on-board energy resources under a large enough time period $T$=120 s. As expected, with the same $T$, more users are covered by increasing the total amount of energy resources. On one hand, as ${E_{{\rm{tot}}}}$ increases, the aerial BS moves closer to the users which have been covered to enjoy a better communication condition. Thus, the covered users remain covered with a decreased association time and the redundant time slots are allocated for other users which have not been covered yet. On the other hand, with an increased ${E_{{\rm{tot}}}}$, the UAV is able to move a longer distance to reach the users which are far away from the CIT. 

\subsection{Designed initial trajectory (DIT)}
In this subsection, we evaluate the benefits of DIT as proposed in Section \uppercase\expandafter{\romannumeral4}. By assuming enough on-board energy ${E_{{\rm{tot}}}} = 2.5 \times {10^4}$J, Fig. 8 compares the achieved coverage probability for six different schemes, i.e., 1) CIT, which corresponds to a scheme using circular trajectory centered at ${{\bf{c}}_{\rm{t}}} = {[750,750]^T}$ with optimized scheduling and association variables; 2) DIT, which uses a fixed designed trajectory and optimized scheduling and association variables. Note that the DIT corresponds to the trajectory generated after one iteration of Algorithm 1 and meets the velocity and acceleration constraints; 3) IA-CIT; 4)IA-DIT; 5) Static UAV with TDMA, where the aerial BS is placed at a location which is 100 meters above ${{\bf{c}}_{\rm{t}}}$ and remains static for the whole time period. In addition, the static aerial BS associates with ground users by the same TDMA scheme, so the scheduling and association variables are optimized; 6) Static UAV with FDMA, where the same static aerial BS as in 5) is used, but the access method changes to FDMA. In other word, each user is associated for the entire $T$ but with a reduced a bandwidth ${B_i} = \frac{B}{M} = 1.25 \times {10^5}$ Hz. \\
\indent As regards the performance observed, we can first conclude that, by exploiting the UAV mobility, a much better coverage performance can be achieved thanks to the reduced communication path loss. Even the CIT scheme which has the worst performance among techniques considering moving aerial BSs achieves an up to 20\% higher coverage probability than static aerial BS using TDMA. More users can be covered by the static aerial BS when FDMA is applied. However, the technique still covers at least 25\% less users than the techniques using moving aerial BSs except CIT when $T$=120 s. As expected, the DIT itself covers a clearly increased number of ground users compared to CIT, and the performance gap between DIT and CIT becomes larger as $T$ increases. This is because DIT moves closer to each of the users and enjoys a better communication channel compared to CIT. Finally, it can be seen that the use of DIT further increases the coverage probability of the proposed iterative algorithm. It is mentionable that, IA-CIT can not achieve 100 \% coverage probability even with large enough $T$. IA-DIT, on the other hand, is able to fill the performance gap and cover all the ground users as long as enough $T$ and ${E_{{\rm{tot}}}}$ is given.\\
\indent It is verified in Fig. 9 that the designed initial trajectory can speed up the convergence. Fig. 9 shows that IA-CIT requires at most 15 iterations to converge while IA-DIT is guaranteed to converge within 9 iterations. This is as expected since the trajectory optimization forces the UAV to move closer to the covered users as shown in the previous section, and DIT has a much reduced distance with the covered users compared to CIT. \\
\begin{figure}[t!]
\centering
\includegraphics[width=0.8\linewidth, height=5.8cm]{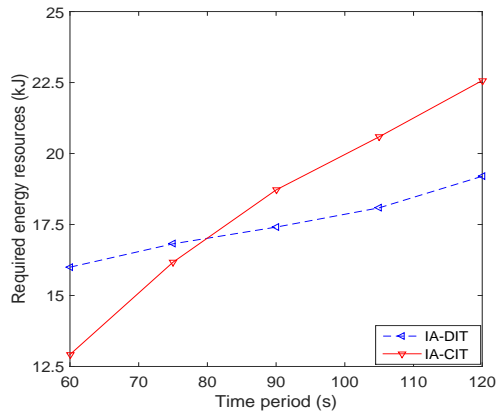}
\caption{Average energy consumption for IA-CIT and IA-DIT, ${E_{{\rm{tot}}}} = 2.5 \times {10^4}$J }
\label{10}
\end{figure}
\indent Fig. 10 further compares the average energy consumption for IA-CIT and IA-DIT, with ${E_{{\rm{tot}}}} = 2.5 \times {10^4}$J. It can be observed that IA-DIT consumes more energy than IA-CIT when we have a short time period. This is because with a limited time slots for association, IA-DIT requires the aerial BS to move faster and change its directions more abruptly compared to IA-CIT. However, the energy consumption of IA-CIT increases more drastically than IA-DIT as $T$ increases. To be specific, the average consumed energy of IA-CIT starts to exceed IA-DIT when $T$=80s, and IA-CIT consumes approximately ${\rm{3.5}} \times {\rm{1}}{{\rm{0}}^{\rm{3}}}$ more energy when $T$=120s.

\subsection{ULI-robust techniques}
\begin{figure}[t!]
\centering
\includegraphics[width=1\linewidth, height=5.8cm]{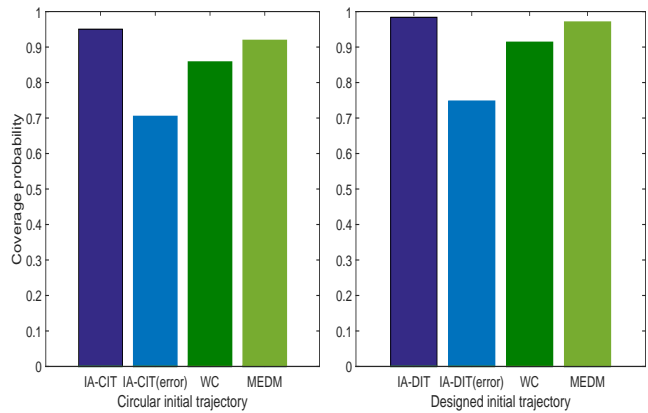}
\caption{Coverage probability with imperfect ULI and Robust techniques, ${E_{{\rm{tot}}}} = 2.5 \times {10^4}$J, $T$=100 s}
\label{11}
\end{figure}
\begin{figure}[t!]
\centering
\includegraphics[width=0.8\linewidth, height=5.8cm]{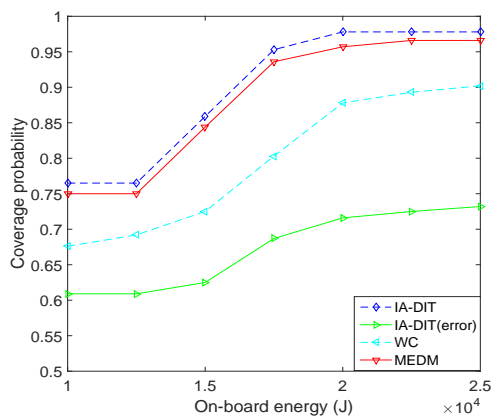}
\caption{Coverage performance of Robust techniques versus various on-board energy ${E_{{\rm{tot}}}}$, $T$=100 s}
\label{12}
\end{figure}
\indent In real scenarios, ULI can only be estimated inaccurately. Therefore, it is meaningful to examine the coverage performance of the proposed IA-CIT and IA-DIT techniques with imperfect ULI. We assume ${E_{{\rm{tot}}}} = 2.5 \times {10^4}$J and $T$=100 s, and the corresponding results are shown in Fig. 11. It can be seen that the performance of both IA-CIT and IA-DIT decreases significantly when introducing imperfect ULI. Take IA-DIT for example, approximately 98\% of total users can be covered when ULI is estimated accurately. However, the coverage probability decreases by 25\% in the existence of inaccurate ULI, since the data transmission process suffers more path loss than expected. It is worth highlighting that the performance loss is greatly compensated when the proposed two robust techniques are applied as shown in Figure. 11. For compensating the performance loss, WC technique guarantees the worst case performance and thus increases the immunity to imperfect ULI. Note that the MEDM technique which provides excessive data to each of the covered users achieves even better coverage probability than WC. When DIT is used, the decreased coverage performance in the existence of inaccurate ULI is almost completely compensated with the use of MEDM. \\
\indent Fig. 12 further shows the coverage performance of the proposed robust techniques versus various on-board energy resources. First note that the achieved coverage probability of IA-DIT increases as more on-board energy is available, which is consistent with the result shown in Fig. 7. In the meanwhile, the coverage performance decreases by more than 20 \% after introducing imperfect ULI. With increased on-board energy, the UAV is able to move closer to the worst case user locations to enjoy better communication condition, so the WC technique is able to cover more ground users in the existence of inaccurate ULI. Similarly, more performance loss can be compensated by MEDM thanks to increased on-board energy, since more excessive data can be provided to the covered users. Also note that an approximately 7\% coverage performance gap remains between WC and MEDM with the change of  ${E_{{\rm{tot}}}}$.

\section{Conclusion}
In this paper, a UAV based aerial BS transmission is considered, where an aerial BS is dispatched for covering a maximum number of ground users before exhausting the on-board energy. An iterative algorithm based on successive convex optimization and block coordinate descent techniques is proposed. The iterative algorithm alternately optimizes the UAV trajectory and user scheduling and association in each iteration. In order to speed up the convergence of the iterative algorithm and achieve a better coverage performance, the initial trajectory is carefully designed so all the ground users have a fair opportunity to be scheduled and associated. In addition, the situation of inaccurate ULI is considered and two different robust techniques are proposed correspondingly. Numerical results verifies that the performance loss due to imperfect ULI can be greatly compensated by the proposed robust techniques.


%





\ifCLASSOPTIONcaptionsoff
  \newpage
\fi



%


\bibliographystyle{IEEEtran}
\bibliography{list}

\begin{thebibliography}{10}
\providecommand{\url}[1]{#1}
\csname url@samestyle\endcsname
\providecommand{\newblock}{\relax}
\providecommand{\bibinfo}[2]{#2}
\providecommand{\BIBentrySTDinterwordspacing}{\spaceskip=0pt\relax}
\providecommand{\BIBentryALTinterwordstretchfactor}{4}
\providecommand{\BIBentryALTinterwordspacing}{\spaceskip=\fontdimen2\font plus
\BIBentryALTinterwordstretchfactor\fontdimen3\font minus
  \fontdimen4\font\relax}
\providecommand{\BIBforeignlanguage}[2]{{%
\expandafter\ifx\csname l@#1\endcsname\relax
\typeout{** WARNING: IEEEtran.bst: No hyphenation pattern has been}%
\typeout{** loaded for the language `#1'. Using the pattern for}%
\typeout{** the default language instead.}%
\else
\language=\csname l@#1\endcsname
\fi
#2}}
\providecommand{\BIBdecl}{\relax}
\BIBdecl

\bibitem{DBLP:journals/corr/abs-1803-00680}
\BIBentryALTinterwordspacing
M.~Mozaffari, W.~Saad, M.~Bennis, Y.~Nam, and M.~Debbah, ``{A Tutorial on UAVs
  for Wireless Networks: Applications, Challenges, and Open Problems},''
  \emph{CoRR}, vol. abs/1803.00680, 2018. [Online]. Available:
  \url{http://arxiv.org/abs/1803.00680}
\BIBentrySTDinterwordspacing

\bibitem{7470933}
Y.~Zeng, R.~Zhang, and T.~J. Lim, ``{Wireless Communications with Unmanned
  Aerial Vehicles: Opportunities and Challenges},'' \emph{IEEE Communications
  Magazine}, vol.~54, no.~5, pp. 36--42, May 2016.

\bibitem{8292783}
E.~Kalantari, I.~Bor-Yaliniz, A.~Yongacoglu, and H.~Yanikomeroglu, ``{User
  Association and Bandwidth Allocation for Terrestrial and Aerial Base Stations
  with Backhaul Considerations},'' in \emph{2017 IEEE 28th Annual International
  Symposium on Personal, Indoor, and Mobile Radio Communications (PIMRC)}, Oct
  2017, pp. 1--6.

\bibitem{7461487}
B.~Galkin, J.~Kibilda, and L.~A. DaSilva, ``{Deployment of UAV-mounted Access
  Points According to Spatial User Locations in Two-tier Cellular Networks},''
  in \emph{2016 Wireless Days (WD)}, March 2016, pp. 1--6.

\bibitem{Namuduri:2013:MAH:2491260.2491265}
K.~Namuduri, Y.~Wan, and M.~Gomathisankaran, ``{Mobile Ad Hoc Networks in the
  Sky: State of the Art, Opportunities, and Challenges},'' in \emph{Proceedings
  of the Second ACM MobiHoc Workshop on Airborne Networks and Communications},
  ser. ANC '13.\hskip 1em plus 0.5em minus 0.4em\relax New York, NY, USA: ACM,
  2013, pp. 25--28.

\bibitem{6863654}
A.~Al-Hourani, S.~Kandeepan, and S.~Lardner, ``{Optimal LAP Altitude for
  Maximum Coverage},'' \emph{IEEE Wireless Communications Letters}, vol.~3,
  no.~6, pp. 569--572, Dec 2014.

\bibitem{7417609}
M.~Mozaffari, W.~Saad, M.~Bennis, and M.~Debbah, ``{Drone Small Cells in the
  Clouds: Design, Deployment and Performance Analysis},'' in \emph{2015 IEEE
  Global Communications Conference (GLOBECOM)}, Dec 2015, pp. 1--6.

\bibitem{7918510}
M.~Alzenad, A.~El-Keyi, F.~Lagum, and H.~Yanikomeroglu, ``{3-D Placement of an
  Unmanned Aerial Vehicle Base Station (UAV-BS) for Energy-Efficient Maximal
  Coverage},'' \emph{IEEE Wireless Communications Letters}, vol.~6, no.~4, pp.
  434--437, Aug 2017.

\bibitem{7510820}
R.~I. Bor-Yaliniz, A.~El-Keyi, and H.~Yanikomeroglu, ``{Efficient 3-D Placement
  of an Aerial Base Station in Next Generation Cellular Networks},'' in
  \emph{2016 IEEE International Conference on Communications (ICC)}, May 2016,
  pp. 1--5.

\bibitem{8587183}
J.~Sun and C.~Masouros, ``Deployment strategies of multiple aerial bss for user
  coverage and power efficiency maximization,'' \emph{IEEE Transactions on
  Communications}, pp. 1--1, 2018.

\bibitem{8329013}
J.~Lyu, Y.~Zeng, and R.~Zhang, ``{UAV-Aided Offloading for Cellular Hotspot},''
  \emph{IEEE Transactions on Wireless Communications}, vol.~17, no.~6, pp.
  3988--4001, June 2018.

\bibitem{7556368}
------, ``Cyclical multiple access in uav-aided communications: A
  throughput-delay tradeoff,'' \emph{IEEE Wireless Communications Letters},
  vol.~5, no.~6, pp. 600--603, Dec 2016.

\bibitem{8304088}
Q.~Wu and R.~Zhang, ``{Delay-constrained throughput maximization in UAV-enabled
  OFDM systems},'' in \emph{2017 23rd Asia-Pacific Conference on Communications
  (APCC)}, Dec 2017, pp. 1--6.

\bibitem{8247211}
Q.~Wu, Y.~Zeng, and R.~Zhang, ``{Joint Trajectory and Communication Design for
  Multi-UAV Enabled Wireless Networks},'' \emph{IEEE Transactions on Wireless
  Communications}, vol.~17, no.~3, pp. 2109--2121, March 2018.

\bibitem{8269064}
J.~Lu, S.~Wan, X.~Chen, and P.~Fan, ``{Energy-Efficient 3D UAV-BS Placement
  versus Mobile Users' Density and Circuit Power},'' in \emph{2017 IEEE
  Globecom Workshops (GC Wkshps)}, Dec 2017, pp. 1--6.

\bibitem{7510870}
M.~Mozaffari, W.~Saad, M.~Bennis, and M.~Debbah, ``{Optimal transport theory
  for power-efficient deployment of unmanned aerial vehicles},'' in \emph{2016
  IEEE International Conference on Communications (ICC)}, May 2016, pp. 1--6.

\bibitem{7192644}
K.~Li, W.~Ni, X.~Wang, R.~P. Liu, S.~S. Kanhere, and S.~Jha,
  ``{Energy-Efficient Cooperative Relaying for Unmanned Aerial Vehicles},''
  \emph{IEEE Transactions on Mobile Computing}, vol.~15, no.~6, pp. 1377--1386,
  June 2016.

\bibitem{6978873}
S.~Kandeepan, K.~Gomez, L.~Reynaud, and T.~Rasheed, ``{Aerial-terrestrial
  Communications: Terrestrial Cooperation and Energy-efficient Transmissions to
  Aerial Base Stations},'' \emph{IEEE Transactions on Aerospace and Electronic
  Systems}, vol.~50, no.~4, pp. 2715--2735, October 2014.

\bibitem{7888557}
Y.~Zeng and R.~Zhang, ``{Energy-Efficient UAV Communication With Trajectory
  Optimization},'' \emph{IEEE Transactions on Wireless Communications},
  vol.~16, no.~6, pp. 3747--3760, June 2017.

\bibitem{8329973}
M.~Hua, Y.~Wang, Z.~Zhang, C.~Li, Y.~Huang, and L.~Yang, ``{Power-Efficient
  Communication in UAV-Aided Wireless Sensor Networks},'' \emph{IEEE
  Communications Letters}, vol.~22, no.~6, pp. 1264--1267, 2018.

\bibitem{7486987}
M.~Mozaffari, W.~Saad, M.~Bennis, and M.~Debbah, ``{Efficient Deployment of
  Multiple Unmanned Aerial Vehicles for Optimal Wireless Coverage},''
  \emph{IEEE Communications Letters}, vol.~20, no.~8, pp. 1647--1650, Aug 2016.

\bibitem{7762053}
J.~Lyu, Y.~Zeng, R.~Zhang, and T.~J. Lim, ``{Placement Optimization of
  UAV-Mounted Mobile Base Stations},'' \emph{IEEE Communications Letters},
  vol.~21, no.~3, pp. 604--607, March 2017.

\bibitem{7366709}
M.~Hong, M.~Razaviyayn, Z.~Luo, and J.~Pang, ``{A Unified Algorithmic Framework
  for Block-Structured Optimization Involving Big Data: With Applications in
  Machine Learning and Signal Processing},'' \emph{IEEE Signal Processing
  Magazine}, vol.~33, no.~1, pp. 57--77, Jan 2016.

\bibitem{8443133}
J.~Zhang, Y.~Zeng, and R.~Zhang, ``{UAV-Enabled Radio Access Network:
  Multi-Mode Communication and Trajectory Design},'' \emph{IEEE Transactions on
  Signal Processing}, vol.~66, no.~20, pp. 5269--5284, Oct 2018.

\bibitem{7572068}
Y.~Zeng, R.~Zhang, and T.~J. Lim, ``{Throughput Maximization for UAV-Enabled
  Mobile Relaying Systems},'' \emph{IEEE Transactions on Communications},
  vol.~64, no.~12, pp. 4983--4996, Dec 2016.

\bibitem{qual}
Qualcomm, ``{LTE Unmanned Aircraft Systems-Trial Report},'' \emph{Qualcomm
  Technologies}, February 2017.

\bibitem{8316986}
D.~Yang, Q.~Wu, Y.~Zeng, and R.~Zhang, ``{Energy Trade-off in Ground-to-UAV
  Communication via Trajectory Design},'' \emph{IEEE Transactions on Vehicular
  Technology}, pp. 1--1, 2018.

\bibitem{Boyd:2004:CO:993483}
S.~Boyd and L.~Vandenberghe, \emph{{Convex Optimization}}.\hskip 1em plus 0.5em
  minus 0.4em\relax New York, NY, USA: Cambridge University Press, 2004.

\bibitem{article}
N.~T. Li, L.~Ting, N.~Adila~Husna, and H.~Husin, ``{GPS Systems Literature:
  Inaccuracy Factors And Effective Solutions},'' \emph{International journal of
  Computer Networks \& Communications}, vol.~8, pp. 123--131, 04 2016.

\end{thebibliography}
%







\end{document}